\documentclass{article}

\usepackage{arxiv}

\usepackage[utf8]{inputenc} 
\usepackage[T1]{fontenc}    
\usepackage{hyperref}       
\usepackage{url}            
\usepackage{booktabs}       
\usepackage{amsfonts}       
\usepackage{nicefrac}       
\usepackage{microtype}      
\usepackage{lipsum}
\usepackage{natbib}

\usepackage{lineno}

\usepackage{graphicx}	
\usepackage{amsmath}	
\usepackage{amssymb}	

\title{Constraining spatial pattern of early activity of comet 67P/C-G with 3D modeling of the MIRO observations}

\author{
Y. Zhao,\thanks{E-mail: zhaoyuhui@pmo.ac.cn}\\
$^{1}$CAS Key Laboratory of Planetary Sciences\\ 
Purple Mountain Observatory, Chinese Academy of Sciences,\\
210008, Nanjing, China\\
$^{2}$Max-Planck-Institut f\"{u}r Sonnensystemforschung,\\
Justus-von-Liebig-Weg 3, 37077 G\"{o}ttingen, Germany\\
$^{3}$CAS Center for Excellence in Comparative Planetology,\\
Chinese Academy of Sciences, China\\
\And
L. Rezac,\thanks{E-mail: rezac@mps.mpg.de}\\
$^{2}$Max-Planck-Institut f\"{u}r Sonnensystemforschung,\\
Justus-von-Liebig-Weg 3, 37077 G\"{o}ttingen, Germany\\
\And
P. Hartogh\\
$^{2}$Max-Planck-Institut f\"{u}r Sonnensystemforschung,\\
Justus-von-Liebig-Weg 3, 37077 G\"{o}ttingen, Germany\\
\And
J. Ji\\
$^{1}$CAS Key Laboratory of Planetary Sciences\\ 
Purple Mountain Observatory, Chinese Academy of Sciences,\\
210008, Nanjing, China\\
$^{3}$CAS Center for Excellence in Comparative Planetology,\\
Chinese Academy of Sciences, China\\
\And
R. Marschall\\
$^{4}$International Space Science Institute,\\
Hallerstrasse 6, CH-3012 Bern, Switzerland\\
\And
and H. U. Keller\\
$^{5}$Institute for Geophysics and Extraterrestrial Physics,\\
TU Braunschweig, 38106 Braunschweig, Germany\\
}

\begin{document}

\maketitle

\begin{abstract}
   Our aim is to investigate early activity (July 2014) of 67P/CG  with 3D coma and radiative transfer modeling of MIRO measurements, accounting for nucleus shape, illumination, and orientation of the comet. We investigate MIRO line shape information for spatial distribution of water activity on the nucleus during the onset of activity. During this period we show that MIRO line shape have enough information to clearly isolate contribution from Hapi and Inhotep independently, and compare it to the nominal case of activity from the entire illuminated surface. We also demonstrate that spectral line shapes differ from the 1D model for different viewing geometries and coma conditions relevant to this study. Specifically, line shapes are somewhat sensitive to the location of the terminator in the coma. At last, fitting the MIRO observations we show that the Imhotep region (possible extended source of H$_{2}$O due to CO$_{2}$ activity) contributes only a small fraction of the total number of water molecules into MIRO beam in the early activity. On the other hand, a strong enhancement of water activity from the Hapi region seems required to fit the MIRO line shapes. This is consistent with earlier Rosetta results. Nevertheless, within the assumption of our coma and surface boundary conditions we cannot get a reasonable fit to all MIRO mapping observations in July 2014, which may illustrate that a more sophisticated coma model or more accurate temperature/velocity distribution is needed.
\end{abstract}

\section{Introduction}
The Microwave Instrument on the Rosetta Orbiter (MIRO) was designed to study the sub-surface heat transport as well as gas activity of the nucleus of 67P/Churyumov-Gerasimenko. Observations of MIRO began on 24 May 2014, and on 6 June, the first water vapor spectral signature from 67P was detected at a heliocentric distance of 3.92~AU \citep{Gulkis:2015}, which was the second largest distance for water detection in a comet after \cite{BMorvan:2010}. During the June-July time interval, Rosetta approached 67P going from a distance of 550,000~km to 973~km, during which the field of view (FOV) of this single-pixel heterodyne spectrometer ranged from about 1200~km to 2~km. Hence during this entire approach phase, the MIRO FOV remained larger than the effective radius of the nucleus of approximately 3.8~km for most of the time, providing an opportunity to study global water production rates. The MIRO observations during the 2014 June-July period are referred to here as the pre-rendezvous observations.

\citet{Gulkis:2015} analyzed early observations acquired by MIRO with a 1D radiative transfer model relying on the assumption of fully homogeneous semi-spherical nucleus (the Haser model). The total H$_2$O production rates were determined from the 557~GHz main isotope water line and monitored during the June-July period, which showed a periodical change related to nucleus rotation. Both, the significant blue shift of the spectral line and the time variations of the production rate indicated a preferential outgassing from the day side of the nucleus. The observations interpreted with the 1D model required a negligible activity from the unilluminated hemisphere, therefore, demonstrating a significant temperature contrast between the day and nightside as well as negligible contribution from extended sources into un-illuminated regions. 

The images acquired by the OSIRIS (Optical, Spectroscopic, and Infrared Remote Imaging System) during August-September 2014 lead to the first detailed characterization of the nucleus shape and yielded the digital shape model \citep{Sierks:2015a}. The nucleus of 67P is characterized by a small and a large lobe connected by a narrow ``neck'' region. This information was used at later observations (August 2014) to suggest that the water outgassing are mainly localized in the neck region \citep{Gulkis:2015,Lee:2015}.  \citet{Bieler:2015} mapped water column density using the Rosetta Orbiter Spectrometer for Iron and Neutral Analysis (ROSINA) data when the comet was at 3.4~AU from the Sun and also showed the strongest activity to be correlated with Rosetta's positions around the neck and north pole of the nucleus on the day side. \citet{Lee:2015} presented results indicating a noticeable spatial and diurnal variation of water outgassing on 67P based on MIRO observations when the spacecraft was already in the ``close nucleus'' orbit ($\approx$100~km) in August 2014. The authors pointed out that the largest water outgassing  during this month occurred near the neck region, and outgassing activities in this region were found to be correlated with the local illumination condition, however, this was not the case for other regions.

MIRO observations during the spacecraft approach phase provided important information on the general mean outgassing, however, these results were limited, because the shape of 67P was not available when Rosetta first arrived at the comet. Hence, the detailed shape was not fully known and could not be taken into account during the first MIRO spectra interpretation. In this work we aim to extend the earlier studies of MIRO measurements from the pre-rendezvous stage, using the 3D nucleus shape model, appropriate illumination conditions on each facet and exact MIRO geometry. Our goal is to 1) constrain the water activity distribution on the nucleus using all the information in the MIRO measured line shape, 2) confirm or rule out possible extended sources of water in the coma from the Imhotep region which showed a strong early activity in CO$_{2}$, and 3) take a first step forward from 1D idealizations used in current MIRO interpretations to begin to identify and study missing physics in the current understanding of the onset of activity of 67P/CG.

In section 2, we provide a summary of the instrument and relevant measurements. In Section 3, the analytical description of the 3D coma model using the shape model and appropriate illumination conditions is described, along with the 3D radiative transfer code LIME \citep{Brinch:2010}, used for calculating rotational populations of water molecules in the coma. In Section 4, we present a synthetic study of sensitivity of MIRO line shape to activity from different areas and for different nucleus orientations, and in section 5, we apply the model to the MIRO measurements. Finally, in section 6 we discuss the results and provide our conclusions.

\section{Observations}
\subsection{MIRO instrument}
The Microwave Instrument for the Rosetta Orbiter (MIRO) acts as a small sub-millimeter telescope, with a parabolic primary dish of 30~cm in diameter. The collected radiation is processed with two heterodyne receivers, one operating at frequency of 190~GHz ($\approx$1.6~mm) and another one sensitive in the range of 562~GHz (0.5~mm), which yields a beam field-of-view of 2.2$\times10^{-3}$~[rad] FWHM (see \citet{Gulkis:2007} for detailed instrument description). Both channels have broadband continuum detector measuring the continuum brightness temperature, and the sub-millimeter band is further processed by the chirp transform spectrometer \citep{Hartogh:1990} providing high spectral resolution for eight molecular transitions. The spectroscopic measurements are done in a frequency switched mode with a shift of $\pm$5~MHz from the nominal frequency every 5 seconds during an integration time. This technique is used to minimize effects of baseline ripple and the possible short term variations in gain drift \citep{Gulkis:2007b}. The radiometric calibration is accomplished by (beam switching) observing hot and cold loads every 30~minutes to determine the gains of the two receivers.

\subsection{MIRO observations: July 2014}
The MIRO instrument registered comet activity mainly from the 557~GHz ortho-H$_{2}^{16}$O transition in early Jun 2014 with estimated production rate Q[H$_{2}$O]$\approx$ 1$\times$10$^{25}$ molecules/sec \citep{Gulkis:2015}. At the time the heliocentric distance was about 3.95~au, well outside the water snow line. In fact all the unresolved nucleus observations, including the ones we aim to discuss in this paper happen at heliocentric distance larger than $\sim$3.6~au. An incomplete log of pre-rendezvous observations made by MIRO are summarized in Table\ref{tbl:log} including a flag indicating to which we applied our 3D model and with what degree of success (which will be discussed later in the paper). 
\begin{table*}
\caption{Log of selected pre-rendezvous MIRO mapping observations}
\begin{center}
{ \hfill{}
\begin{tabular}{p{4cm} c c c c c c }\hline\hline\noalign{\vspace{1ex}}
Date$^{a}$ & Flag$^{b}$ &N$^{c}$ & Integration$^{d}$ & $\langle r_{h} \rangle^{e}$ & $\langle \Delta \rangle^{f}$ & $\langle \phi \rangle^{g}$\\
						(UTC) & & & (min) & (AU) & (km) & (deg) \\
					\hline \noalign{\vspace{1ex}}
2014-07-05T02:41 & 0 &5$\times$5 &  26  & 3.78 &  3.63$\times$10$^{4}$ & 24 \\
2014-07-05T15:44 & 0 &3$\times$3  &  66  & 3.78 &  3.43$\times$10$^{4}$ & 23 \\
2014-07-11T22:36 & 1 &3$\times$3 &  38  & 3.78 &  1.58$\times$10$^{4}$ & 23 \\
2014-07-12T00:49 & 1 &3$\times$3 &  110  & 3.78 & 1.50$\times$10$^{4}$ & 23 \\
2014-07-19T05:01 & 1 &3$\times$3 &  27  & 3.70 &  6.60$\times$10$^{3}$ & 8 \\
2014-07-19T11:44 & 1-2 &5$\times$5 &  12  & 3.69 &  6.40$\times$10$^{3}$ & 8 \\			
2014-07-19T17:43 & 1-2 &6$\times$5 &  10  & 3.69 &  6.22$\times$10$^{3}$ & 7 \\	
2014-07-19T23:56 & 1-2 &3$\times$3 &  37  & 3.69 &  6.05$\times$10$^{3}$ & 7 \\	
2014-07-27T17:27 & 2 &7$\times$7 &  5   & 3.65 &  2.23$\times$10$^{3}$ & 1 \\
\hline
\end{tabular}}
\hfill{}
\label{tbl:log}
\end{center}
{\small \textbf{Notes.} $^{(a)}$ Approximate time of mid-point observation in a format yyyy-mm-ddThr:min. $^{(b)}$ Flag=0 we did not use this data, 1 we applied the 3D modeling in this work with success, 2 we applied the 3D modeling, however, fitting is not reliable due to other issues and future work is needed with more detailed model of geometry, coma and/or continuum model. $^{(c)}$ Number of RA and DEC points in the map.  $^{(d)}$ Average on-source integration time per point. $^{(e)}$ Average heliocentric distance for the mapping observation.  $^{f}$ MIRO cometocentric distance, and  $^{(g)}$ average phase angle.	}
\end{table*}

\section{Modeling}
\label{section:models}
The interpretation of MIRO observations requires a 3D radiative transfer model capable accounting for non-LTE effects in molecular excitation. The radiative transfer model requires 3D information on water molecules, temperature and velocity distribution, along with spectroscopic parameters. In addition, we need a fast and reliable synthetic beam averaged spectra generator relying on the output populations of the 3D transfer code, along with precise MIRO pointing information. 

We adopted the LIME package \citep{Brinch:2010} (detailed later) for calculating the rotational populations of ortho-water in full 3D, and develop a routine to evaluate MIRO beam averaged radiance using precise pointing information from SPICE kernels \citep{Acton:1996}. These codes in turn require as an input a shape model (serving as a boundary condition), and also a coma characterized by water density, temperature, and expansion velocity covering the relevant computational domain. 


\subsection{The shape model}
\begin{figure}
	\includegraphics[width=\columnwidth]{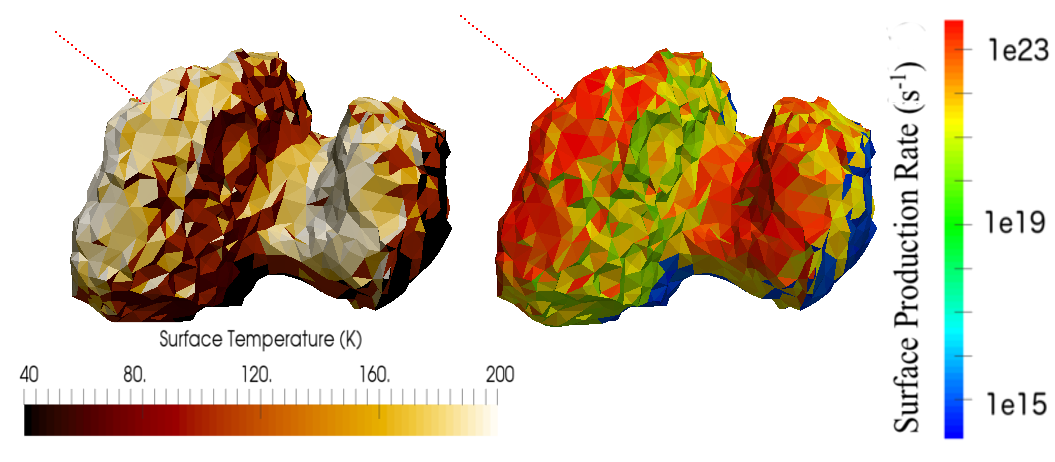}
    \caption{Distribution of surface temperature (left) and facet production rate (right) $Q_{f}$ on CG's surface at epoch UTC 2014-07-12T10:18:32, dashed line in red indicate solar incidence direction. The production rates for each facet, which is the flux multiplying facet area. It is calculated from model A \citet{Keller:2015} and scaled to a total water production rate of $1.02\times10^{26} s^{-1}$.}
    \label{fig:surfaceparameter}
\end{figure}

In this work we adopt nucleus shape model (SHAP7) \citep{Preusker:2017},  but modified by Meshlab\footnote{\url{http://www.meshlab.net/}} to a resolution of 3000 facets. The average distance between two neighboring vertexes is 109.8~m and the average area of the facets for our shape model is 16263 m$^2$. The low spatial resolution of the nucleus for the early observation is adequate since the projected MIRO beam size varies from more than 30 to about 2~km FWHM during this period. Our requirement is only to be able to clearly distinguish the main features of the shape, such as the Imhotep, Hapi (``neck''), or small lobe region. These ``general'' morphological units are used for generating different models of varying sources of water activity. The degraded resolution nucleus shape with the effects of illumination on surface temperature and water production rates for each facet is shown in Fig.~\ref{fig:surfaceparameter}.

\subsection{Coma model} 
\label{sec:model}
The coma generation is the most challenging part because fitting observations may require testing of several tens of full 3D coma models for each point during the MIRO mapping observations. Hence, there is a need for fast evaluations of the coma model. It should be noted that our goal in this work is not to develop a first principles physics, self-consistent coma model. The aim is to have a coma model reflecting the main characteristics consistent with the comet shape, illumination and orientation of the nucleus that at the end fit the MIRO line shapes. The complex, self-consistent model are yet to be developed guided by the knowledge gained from the simpler ones, such as the one presented here.


We adopt a water density model driven purely by insolation impinging on the facets of the digital shape as used in \citet{Keller:2015} (model A) (see Fig.~\ref{fig:surfaceparameter}). A constant production rate of $1\times 10^{12}~m^{-2}s^{-1}$ is assumed for the night side (unilluminated facets), but otherwise the day and night side coma follows the same calculations determining the radial water density profile. It is also important to note that in the methodology and sensitivity analysis sections (sec.~\ref{sec:validation}, \ref{sec:nonLTE}, and \ref{section:numerical}), we scale down the production rate of each illuminated facet by a factor of 26 in order to have the total surface production rate at a magnitude of about $1\times 10^{26}$. In section \ref{sec:application} the production rates are scaled differently in order to match actual MIRO measurements, as will be discussed later. 

For the coma grid we use a log scale to sample spheres centered at 67P's body center with different radii ranging from 0.45 km to 700 km. On each sphere we select $100\times100$ points in spherical coordinates and remove those inside the shape of the nucleus.

As Fig.~\ref{fig:vapour} illustrates the number density $n(z)$ at grid point located at distance $z$ from the facet fulfills the following formula: 
 \begin{figure}
	\includegraphics[width=\columnwidth]{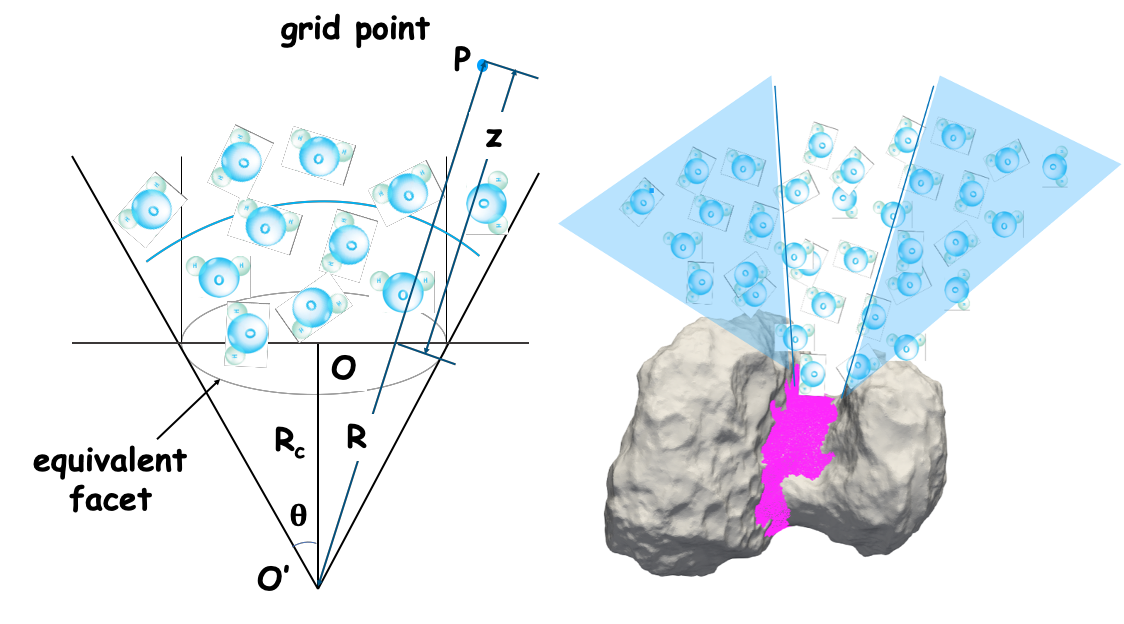}
    \caption{Left: Sketch of water molecules produced by an equivalent facet. A grid point $P$ in the coma is represented in blue, while $z$ and $R$ are surface distance and equivalent radius distance, respectively. $\theta$ is the opening angle in our simulation. $O$ and $O'$ are the facet center and the apex of the cone respectively, $R_c$ is the distance between these two points. Right: Sketch of coma structure produced by the Hapi region as a result of combined effects from opening angle and morphology, which block certain angles in the cone leading to smaller coma density (shaded blue region). }
    \label{fig:vapour}
\end{figure}
\begin{equation}
Q_{f}dt = S_{sphere}(z)n(z){V_{exp}(z)}dt,
\end{equation}
where $Q_{f}=S\times Z$, and $S$ and $Z$ are facet area and production rate $(molecules/m^{2}/s)$ respectively. $S_{sphere}(z)$ is the area of the cross section of the sphere in the cone at surface distance $z$. The opening angle $\theta$, is employed to describe the solid angle $\Omega$ of the surface outgassing with the relationship $\Omega=2\pi(1-\cos\theta)$. Then we have
\begin{equation}
S_{sphere}(z) =\Omega R^2 = 2\pi(1-\cos\theta) R^2.
\end{equation}
$R$ is the radius of the reference sphere centered at $O'$, where  $O'$ is the apex of the cone and could be considered as an equivalent source point with total production rate of $Q_f$, as shown in Fig.~\ref{fig:vapour}. Suppose a circle with the same area as the triangle facet has a radius of $r$, we have the height of the cone as following:
\begin{equation}
R_c = |\vec{OO'}| =\frac{\sqrt[]{\frac{S}{\pi}}}{\tan\theta}.
\end{equation}
where $O$ is the facet center. $\vec{OO'}$ points along the direction of the facet's normal vector. For an arbitrary grid point $P$ in the coma model, we have:
\begin{equation}
R = |\vec{O'P}|,
\end{equation}
surface distance z of a grid point $P$ is defined as the distance between $P$ and the intersection of $\vec{O'P}$ with the facet, as shown in Fig.~\ref{fig:vapour}.

Substitute $(4)$ into $(2)$ and then $(2)$ into $(1)$, taking the finite lifetime against photo-dissociation into consideration, we arive to an expression to determine number density at surface distance z:
\begin{equation}
n=\frac{Q_{f}}{2\pi R^2(1-\cos\theta)V_{exp}}\times e^{(-(10^{-5}/r_{h}^2)\times \frac{z}{v_{exp}})}
\end{equation}
where $10^{-5}$ is the inverse of lifetime of water molecules against photo-dissociation at 1~au scaled by appropriate heliocentric distance, $r_{h}$. 

When $\theta = 180^\circ$, a purely spherical (Haser model) is presented (but nonphysical for an individual facet). For  $\theta = 90^\circ$, the hemispherical Haser limit is recovered above a given facet. 

In addition, we consider the effects of a given facet's surrounding morphology. For this bi-lobed nucleus, Hapi is a special region showing a highly concave structure (right panel in Fig. \ref{fig:vapour}). The cliffs from two lobes prevent some molecules from escaping such that the coma produced by Hapi is diluted (due to an assumption of an absorbing boundary).This 'shadowing' effect is accounted for during test for the visibility of the facets from any given coma sampling point.

The temperature $T$ and velocity $V_{exp}$ are based on the analytical formulas used with reasonable success in the 1D analysis of MIRO measurement in \citet{Gulkis:2015}, \citet{Lee:2015} and \citet{Biver:2015}.  Explicitly, at a given surface distance $z$:
\begin{equation}
V_{exp} = \tanh(z/2400.0)\times v_{m}, \newline
\end{equation}
the direction of velocity is considered to be along the vector of $\vec{O'P}$. For the early observation case, $v_{m}$ has the following values:
\begin{equation}
v_{m}=\left\{
\begin{array}{rcl}
750 m/s     &      & {T_{f} > 250 K}\\
750-2.5\times(250-T_{f}) m/s       &      & {T_{f} <= 250 K}
\end{array} \right.
\end{equation}
Where $T_{f}$ is equilibrium surface temperature and calculated with moadel A in \cite{Keller:2015}. The temperature of the coma at surface distance $z$ is determined by 
\begin{equation}
T = 30.0-2.5\times(30.0-T_{f})(\frac{1}{z})^{0.3} \newline
\end{equation}
Since similar temperature and velocity parametrization have been used in 1D models of \citet{Gulkis:2015,Biver:2015}, they are sufficient for our purposes as well. The main parameters, T$_{f}$ and v$_{m}$, are at the end estimated from the MIRO line width and line shift, in order to give reasonable LOS coma profiles (discussed in section~\ref{sec:application}). It turns out that for these MIRO unresolved nucleus data, even a properly chosen constant profiles would be sufficient for our goals from these early observations.

For any given coma sampling point we evaluate the water density by adding up the contribution of all visible facets. The resultant temperature and velocity for a given coma grid point are computed as a weighted average from all visible facets, where weights are given by the water density arriving from the different surface facets. 

\subsection{Coma model validation}
\label{sec:validation}
\begin{figure}
	\includegraphics[width=\columnwidth]{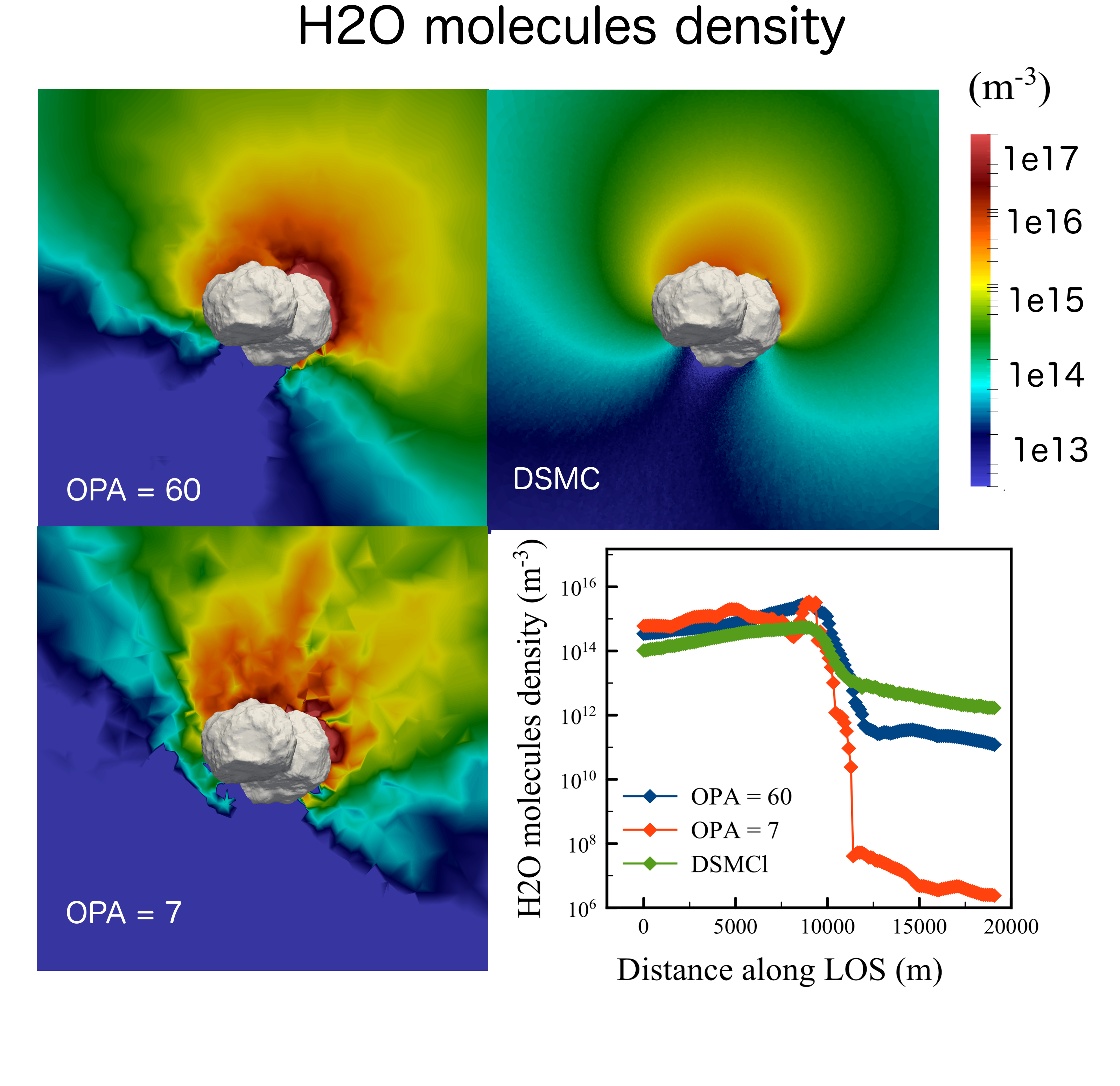}
    \caption{Comparison of the water density coma calculated from our semi-analytical model for two opening angle values (upper and lower left panels) and DSMC (upper right). The figures are slices centered at nucleus' center and with normal direction along x axis pointing to the viewer. The coma cover a domain of about 10~km. The line plots in the lower right panel compare extracted density profiles from the analytical model with opening angle (OPA) 60 and 7 deg, and from the DSMC simulations, along a random pointing direction (single pencil beam) at a tangent distance of 3~km.}
 \label{fig:com_dsmc}
\end{figure}

In our nominal coma model, we use an opening angle of 60$^{\circ}$ (see comparisons and discussion on different opening angles in appendix Fig.~\ref{fig:opa60vs90}). First, we compare our coma model water density distribution to a 3D DSMC simulation \citet{Marschall:2016} in Fig.~\ref{fig:com_dsmc} for heliocentric distance of 3.74~au (Q[H$_{2}$O]$\sim$~10$^{26}$). The images shows slices centered at 67P's center with normal directions along x axis pointing to the viewer. In this qualitative comparison we find that the general coma structure is well reproduced with $\theta=60^{\circ}$. Notably, we also see that the terminator location along a given line of sight (randomly selected in the bottom right panel) is properly captured by the model. This gives us confidence that the shape and illumination of facets are properly implemented. The difference in absolute number density in Fig. 3 is not much of a concern as the production rate will be at the end scaled to match the MIRO line shapes.

\begin{figure}
	\includegraphics[width=\columnwidth]{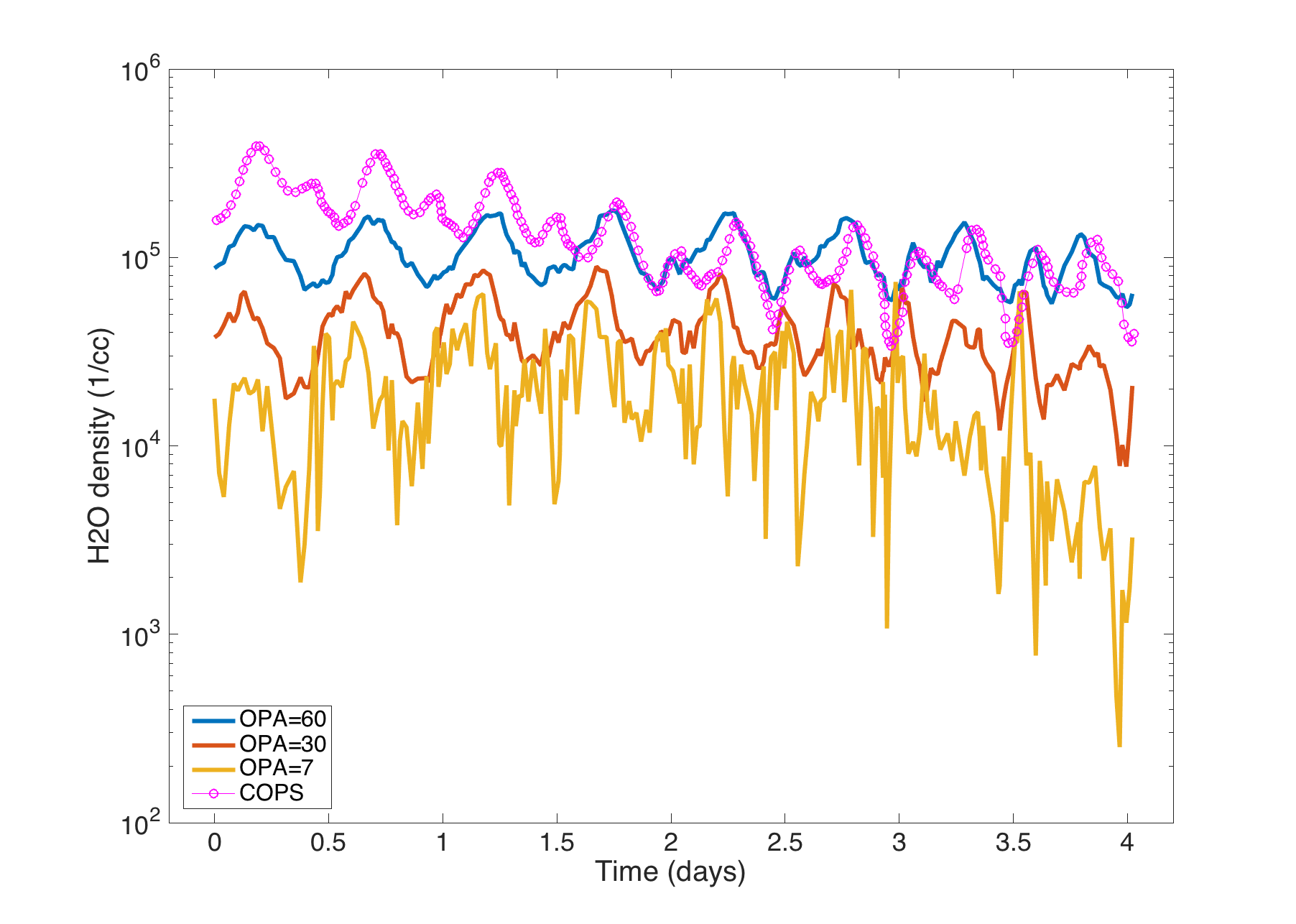}
    \caption{Comparison of water molecules density change at Rosetta's location for days from Sep. 12, 2014, from ROSINA/COPS and our model. Nucleus' rotation states and observing geometries are calculated from SPICE kernels. Line curves in blue, red and orange show results obtained by the coma model with opening angle 7$^\circ$, 30$^\circ$, and 60$^\circ$. The fact that the different modeled curves are offset from one another is just an effect of the density distribution difference due to different opening angle; the total production rate is  $1.02\times10^{26}.$} 
    \label{fig:com_bieler}
\end{figure}

In the second step to validate the coma model, we investigated how well it can capture the diurnal variations due to nucleus rotation. For this we used the published ROSINA/COPS time series data, as shown in Fig.~\ref{fig:com_bieler}. This figure shows the water density variations over a four day period obtained by the ROSINA COPS data (magenta circles) from \citet{Bieler:2015b}, and the simulated water density variations for the same conditions using our nominal coma model. The different model curves correspond to different opening angles to see whether some of the smaller variations observed in COPS timeseries can be better explained with smaller angles, which does not seem to be the case.

From these figures, we conclude that our coma model using the nominal opening angle $\theta=60^{\circ}$ properly captures the main effects due to 3D shape, illumination and variations due to nucleus rotation. 


\subsection{3D non-LTE and beam synthesis model}
\label{sec:nonLTE}
With the 3D distribution of $T$, $V_{exp}$, and $n$ obtained from the coma model described above, we are in position to calculate the non-LTE rotational populations of the ortho-H$_{2}$O. This is accomplished by the LIME package\footnote{\url{https://github.com/lime-rt}}, which is a 3D molecular excitation and radiation transfer code for far-infrared and (sub-)millimeter wavelengths \citep{Brinch:2010}. We use the stable version 1.7, modified to include the nucleus shape model as the lower boundary condition. The code is capable of handling multiple collision partners (water-water and water-electrons), and solar absorption effects in line excitation are also included via the so called g-factors. In this work we use the same ortho-H$_{2}$O non-LTE model with seven levels and nine transitions as used in previous work \citep{Gulkis:2015,Lee:2015}. In our nominal model, following the formalism as in \citep{Lee:2011}, the water-electron collisional excitation may also play a role at larger distance from the nucleus. Nevertheless, the measurements of electron density around 67P during an early activity (July-August 2014) \citep{Edberg:2015} is too low to strongly influence the simulated line amplitudes. 

Once the LIME code outputs populations, we can calculate beam averaged radiance for the precise MIRO pointing as derived from the SPICE kernels for a given time. Unless otherwise specified the time is the middle of the integration duration for a specific pointing. The beam synthesis calculations assume a Gaussian shape sampled such that the projected geometrical step between rings of a constant response is about 200~m (at the comet distance). Each ring is sampled with up to 200 points (depending on the angular distance from the center of the beam), and for each point a pencil beam radiance is calculated. The total radiance is accumulated as an weighted average accounting for the solid angle of each point and the appropriate Gaussian response.

\begin{figure}
	\includegraphics[width=\columnwidth]{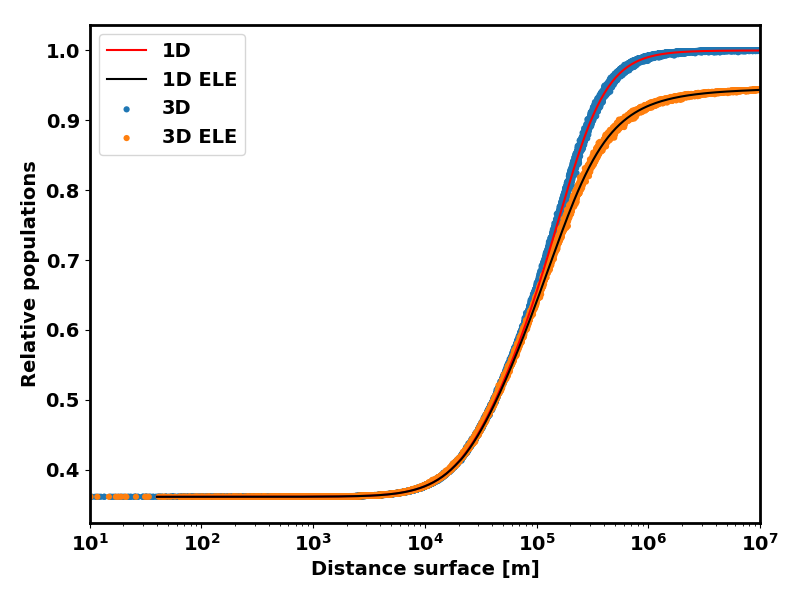}
    \caption{Comparisons of o-H$_{2}$O relative population for the 1$_{01}$ level (ground state) using the 1D and 3D non-LTE calculations. The 1D results are shown as solid lines; red and black corresponding to runs with/without electrons respectively. The blue and orange circles are results of the 3D LIME calculations (with/without electrons respectively as labeled). The presented results are for a spherical coma approximation to retain simplicity of interpretation.}
    \label{fig:pops1d3dali}
\end{figure}

In order to validate accuracy of the 3D populations distribution obtained by LIME, as well as our beam synthesis code we compare them in the limit of spherical symmetry to 1D calculations, used for previous MIRO analysis \citep{Marshall:2017}. The presented calculations are done for a total production rate of 10$^{26}$ molec/sec, with constant temperature of 70~K and expansion velocity of 500~m/s. We choose a little higher production rate to demonstrate the opacity effects in the spectral line formation, both for non-LTE calculations and beam averaged radiances. Both, the 1D and 3D calculations include full radiative transfer to take into account optical depth effect in deriving the mean intensity in the non-LTE. We also note that the purely spherical 3D calculation are done without including the nucleus shape, instead a sphere of 2~km radius is assumed for consistency. First, in Fig.~\ref{fig:pops1d3dali} we show the relative populations of the ground state level, 1$_{01}$, where the largest fraction of molecules resides. The red thin solid line and blue circles compare 1D and 3D results, for a case excluding the excitation by electrons. The solid black line and orange circles compare 1D versus 3D when the electron collisions are included in the simulation (the maximum electron temperature was set to 2000~K). In summary, the 1D and 3D non-LTE populations show excellent agreement, even for multiple-collisional partners, although, in real application the electron excitation will be neglected due to their small density during the early phase. 

\begin{figure}
	\includegraphics[width=\columnwidth]{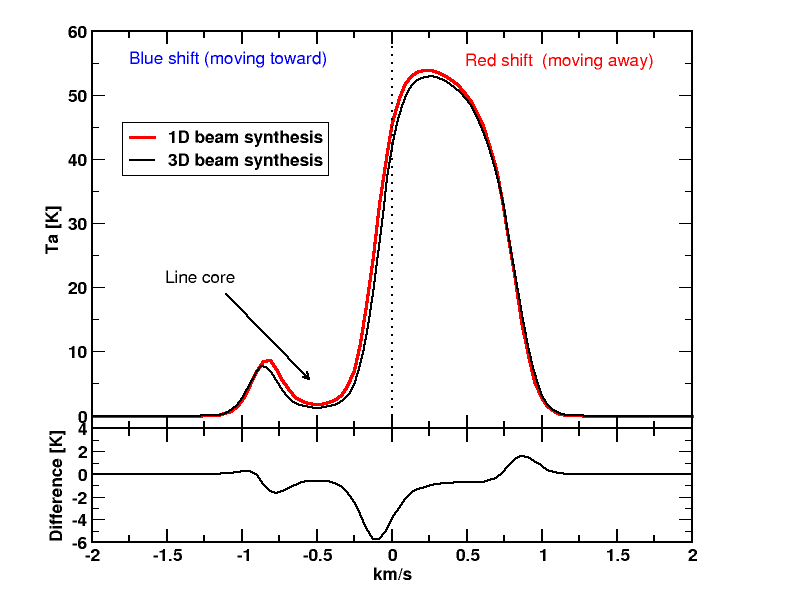}
    \caption{Synthetic beam averaged line shapes simulating MIRO observations from a distance of 15000~km comparing the 1D and 3D beam synthesis using the atmosphere and populations shown in Fig.~\ref{fig:pops1d3dali}. We also annotate the figure indicating the blue and red shifted regions of the spectra (with respect to the observer), and we also note the position of the line core in this example. The bottom panel of the figure shows differences between 1D and 3D simulations.}
    \label{fig:beams1d3d}
\end{figure}

Next we compare the beam averaged spectra derived from the 1D and 3D populations discussed above, as shown in Fig.~\ref{fig:beams1d3d}. The geometry is such that MIRO is pointing at the center of the ``comet'' from a distance of 15000~km (typical distance of early observations). The comparison is very encouraging showing difference of less than 1.5~K in the positive and negative peaks of the line, as well as in the self-absorbed core of the line around -0.5~km/s. The largest discrepancy appears in the line segment with a strong gradient at around -0.1~km/s. This difference is most likely governed by the nature of the 3D beam synthesis rather than due to differences in the populations. The difficulty arises in sampling the 3D grid when generating the LOS pencil beams and projecting the velocity along the ray.

\section{Numerical study of regional outgassing}
\label{section:numerical}
Having validated the 3D non-LTE transfer and line synthesis code, we are now in the position to investigate sensitivity of unresolved MIRO measurements to different distributions of water activity on surface. The aim of this section is to visualize how the MIRO line shapes differ when various surface areas are active. Although, these simulations are not yet designed to provide perfect match to the measurements (only the sensitivity), we do use a real MIRO geometry and pointing relevant for later applications.


\subsection{Effects of regional activity}
\begin{figure}
	\includegraphics[width=\columnwidth]{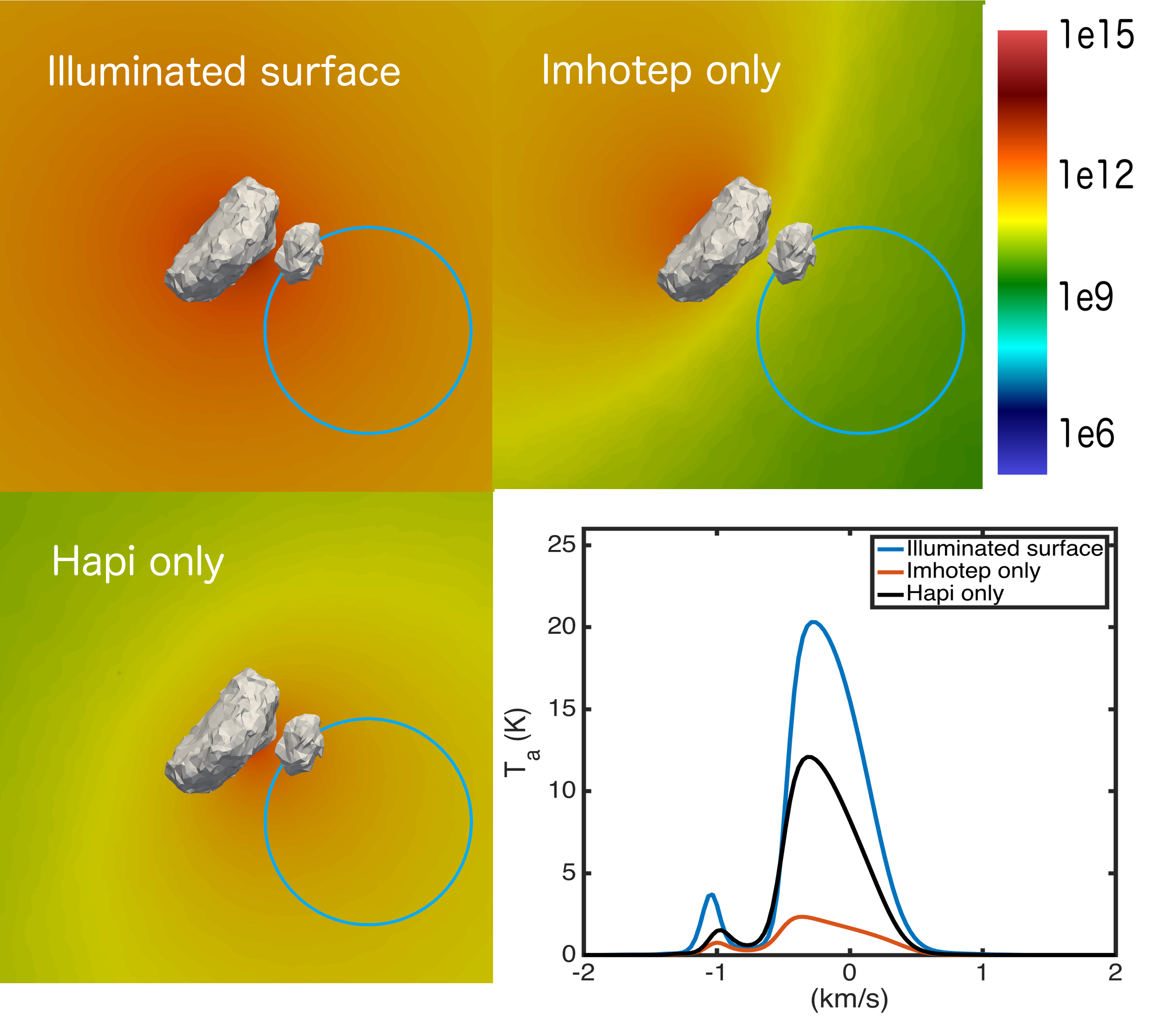}
    \caption{The upper left, upper right and lower left panels show the distribution of water column density [cm$^{-2}$] over a large portion of the coma (more than 60~km) when considering different source regions, as labeled. For Hapi only and Imhotep only models, the facet production rate in the specific region is amplified by a factor of 3 and set to $1\times10^{12}$ in the other parts of the surface. The MIRO beam is also displayed as the blue circle giving the observational geometry context (which is the same for all three panels), and corresponds to the first point of the map on July 12, 2014 (see Fig.~\ref{fig:july12map} for context). The nucleus is also overlayed on the images but it is significantly enlarged (scaled by factor 5) to be visible.}
    \label{fig:regions}
\end{figure}

First, we set out to investigate how much difference in the MIRO line shapes can we expect for different observing geometries when taking into consideration of activity only from distinct nucleus regions, namely Hapi (neck), Imhotep \citep{Maary:2015}, and also the entire illuminated surface. To simulate the regional activity, relevant facets of a given region have their nominal production rates (determined by given illumination conditions) amplified by a factor of 3, while the production rate in all other regions is set to $1\times10^{12}$. The structure of the near nucleus coma is shown in detail for the three cases (Hapi, Imhotep, and illumination) in Fig.~\ref{fig:regions} from the point of view of column density, along with the simulated MIRO lines (bottom right panel) from a distance of 15000~km. The MIRO beam pointing is the same for all the shown cases, but it is not centered at the image center, but offset by about 20~km from the nucleus, as derived from SPICE kernels.

The coma shown in Fig~\ref{fig:regions} indicates clear differences in the structure for the different models. The activity of the entire illuminated surface yield a very homogeneous inner coma from the point of view of column density, while the activity from only the Imhotep region results in a strong heterogeneous distribution, with large high/low contrast. The Hapi only activity, as expected, shows a narrow fan-like activity, with a large fraction of the coma with low density. The differences in the outgassing pattern give rise to a different distribution of number density (and projected velocity) along the MIRO LOS, resulting in strongly different line shapes. The line amplitudes, as well as line areas differ strongly (as the number densities are different), but also the line shapes (contrast red and blue curves). 

\subsection{Effects of illumination/rotation}
\begin{figure}
	\includegraphics[width=\columnwidth]{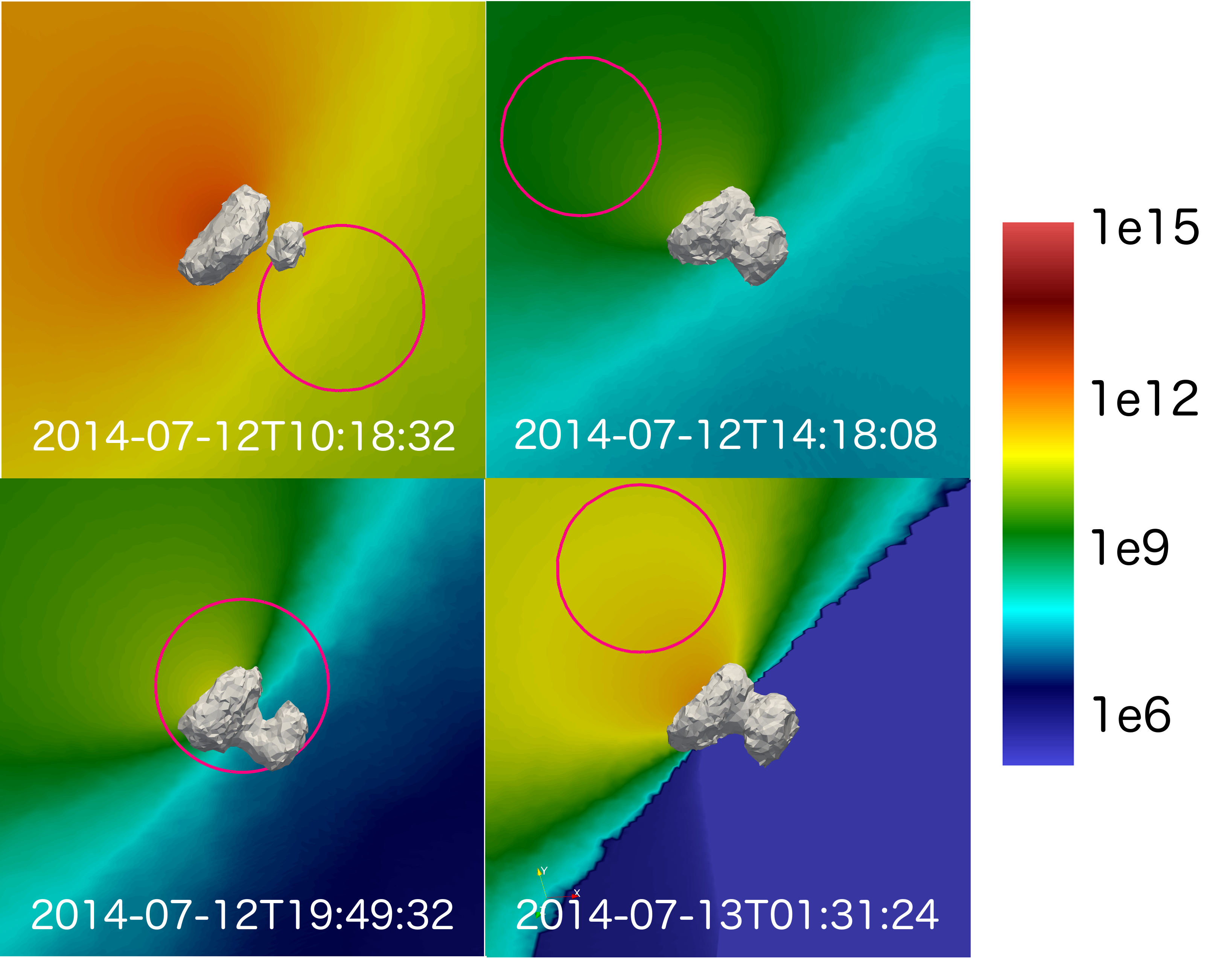}
    \caption{Four panels showing column density [cm$^{-2}$] over a large domain of the coma with the MIRO beam over-plotted as the red circle. The activity is assumed to come only from the Imhotep region to demonstrate the noticeable changes in illumination due to nucleus rotation, therefore the nucleus coma structure. As the nucleus rotates, Imhotep is well illuminated in the upper left, but almost in the night side in upper right and lower left panels. The varying illumination conditions account for the variations in column densities plotted in color for each panel. The panels correspond to four different epochs as shown in labels and are shown as pointings 1,3,6,9 in Fig.~\ref{fig:july12map}). The nucleus shape is scaled by factor of 5 to be visible.}
    \label{fig:imhotep}
\end{figure}

\begin{figure}
	\includegraphics[width=\columnwidth]{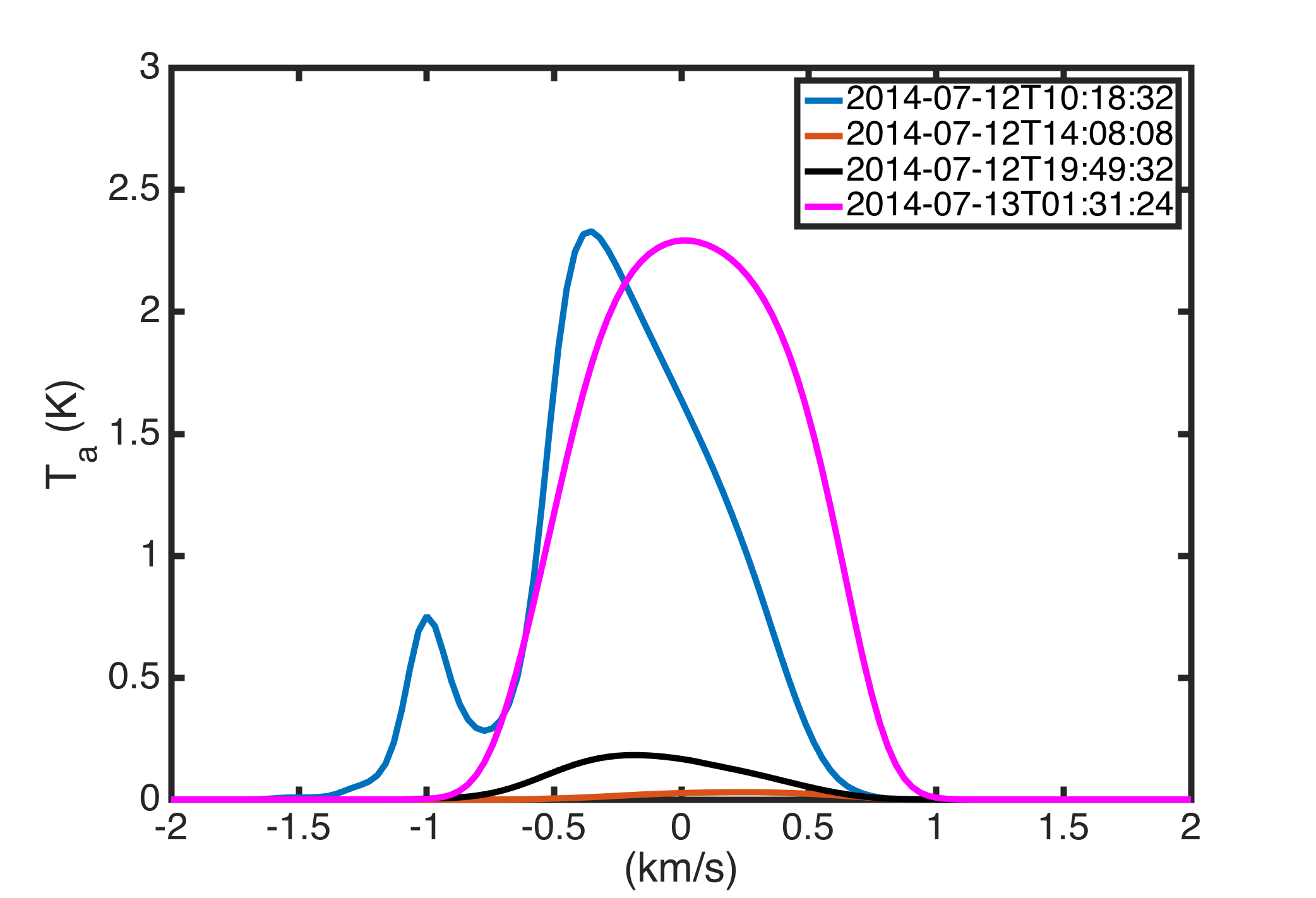}
    \caption{Four spectral lines obtained for the coma models presented in Fig.~\ref{fig:imhotep}, clearly showing the variation of the Imhotep region contribution to the overall MIRO line shape and amplitude. In the extreme case when the Imhotep region is facing away from the MIRO beam, the contribution is negligible (red curve).}
    \label{fig:line_imhotep}
\end{figure}

Next, we will illustrate how effects of nucleus rotation and different observing geometries also produce detectable changes in the spectral line shapes. Each of the four panels in Fig.~\ref{fig:imhotep} corresponds to a different time of MIRO pointing during July 12, 2014, and hence to different rotation phases of the nucleus. The panels show a water column density as color images covering a domain of more than 60~km, with the MIRO beam superimposed as the red circle to give visual context of the overall coma structure. The comet shape is also plotted for reference, but it is again enlarged by factor 5 to be visible. The coma model in this example assumes that only the Imhotep region is active (facets in the Imhotep region are scaled up by a factor of 3 relative to the nominal illuminated model for which the total production rate is Q$\sim$1$\times 10^{26}$, while in the other regions the production rate is set to be $1\times 10^{12}$), in which case the simulated MIRO lines would look like as shown in Fig.~\ref{fig:line_imhotep}.
The simulations in Fig~\ref{fig:imhotep} and \ref{fig:line_imhotep} convey a clear message that  if a particular region, as distinct as Imhotep (same is true for the Hapi region) are active, we would expect to see a rather strong variations in the spectral line shapes during the mapping observations by the MIRO instrument.

\subsection{Effects of observational geometry}
\begin{figure}
	\includegraphics[width=\columnwidth]{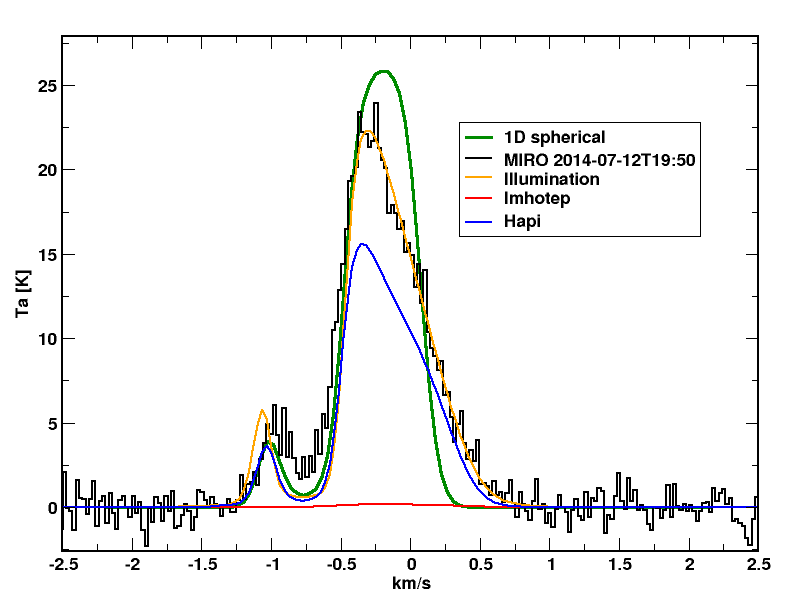}
    \caption{Simulated spectral lines produced from the four different models (as labeled), including 1D simulations, with measured MIRO line also shown. This plot is for the sixth map point in July 12, 2014 (see appendix Fig~\ref{fig:july12map}). }
    \label{fig:lines31}
\end{figure}

\begin{figure}
	\includegraphics[width=\columnwidth]{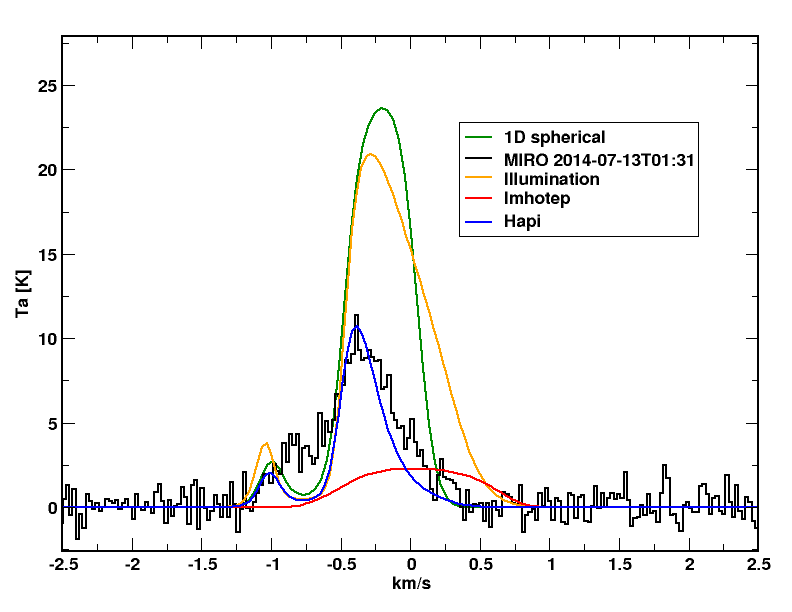}
    \caption{Simulated spectral lines produced from the four different models (as labeled), including 1D simulations, with measured MIRO line also shown. This plot is for the ninth map point in July 12, 2014 (see appendix Fig~\ref{fig:july12map}).}
    \label{fig:lines41}
\end{figure}

In the following comparison, we contrast the MIRO line shapes for two different pointing geometries (hence also different illumination), for the three coma models already discussed (Hapi, Imhotep, and nominal illumination), however, we also add the 1D spherical model results. The MIRO spectra are also shown as the black noisy line in the Figs.~\ref{fig:lines31} and \ref{fig:lines41} for reference. At this point, we are still not aiming to match the MIRO line amplitudes, instead we contrast the different line shapes from the different models (and the MIRO measurement).

 We note that the 1D simulation does not depend on illumination conditions, only the radial offset from nucleus plays a role. Because of this, the 1D synthetic spectral line code is made to take into account only a half of the LOS (closer to the observer) to implicitly introduce a ``night-side''. This issue was already noted in the early investigation \citep{Gulkis:2015}, and the same approach was taken in order to match the measured line widths, which indicate a strong day/night density dichotomy. We note that in these figures we are not trying to reproduce measured line amplitudes (which is done in the following section). Therefore, we still rely on our nominal models as discussed so far, and the focus is placed on understanding how the outgassing from different source regions produces variations in the line shapes, and how they differ from the measured ones (shown in the figures). The aim is to get a qualitative insight whether the MIRO measurements (particular line shape) can be a combination of several regional outgassing regions, or rather if we can deduce their relative contribution in the process of line formation. This reflects not only in water density, but other parameters, such as temperature or expansion velocity should be adjusted. The comparisons are shown in Fig.~\ref{fig:lines31} and \ref{fig:lines41}.

In both figures we see that the 1D calculations provide very weak match to the measured line shape in the region -0.3 to about 0.5~km/s, but they do provide the self-absorbed line core at about -0.7~km/s (which is pronounced in Fig.~\ref{fig:lines31}). The calculated line shapes corresponding to the entire illuminated surface outgassing (orange curves) show a better agreement of the ``red'' wing slope, although the nominal model density would produce slightly large line widths. The ``Hapi only'' case of outgassing shows a small deviation from the full illumination line shape, but it does reproduce both, the ``red'' and ``blue'' portion of the line shape very well. Finally, the ``Imhotep'' region, which happens to  be oriented away from the observer such that we see very little water number density in the beam. Overall, these results suggests that a combination of illumination and enhanced activity in the Hapi region may produce the correct line shape and amplitude.

The second geometry, Fig.~\ref{fig:lines41}, yields a smaller measured line amplitude as the pointing is further away from the nucleus (Fig.~\ref{fig:july12map}), and as a result the line core is also only weakly self-absorbed (but still clearly blue shifted).  In this geometry, the nominal illumination case shows a rather weak match to the measured line shape. On the other hand, the ``Hapi`` only outgasing in this nucleus orientation does a good job reproducing the blue and red wings slopes as well as the red wing line amplitude, but the line width in this case is little too narrow near the base of the spectral line. Therefore, just as pointed out previously, combining the illumination and Hapi activity can perhaps yield a better fitting of the measured line shapes, which is the approach we take in the next section.


\section{Application to MIRO measurements}
\label{sec:application}
Finally, after understanding the line shape sensitivity we are in the position of trying to match the observed line amplitudes and line shape for all the points in the MIRO map performed in July 2014. As indicated in Table \ref{tbl:log}, we do not analyze first two events in the context of this paper as they are too noisy (flag=0 in the table). Events analyzed with reasonable success are indicated with flag=1, and we discuss them in detail in this section. 

First, we analyze the map from July 11, 2014. In order to match the line amplitude and line shape, we made the following adjustments to the production rates. At each illuminated facet, we scale the production rate by 0.1, but the Hapi region is scaled only by 0.5. In previous section we used ``Hapi only'' which was scaled up by factor 3 (which we will call (Hapi$^{0}$)), while in this section we use Hapi only region scale as 0.5*3, which we refer to as (Hapi$^{+}$). We make this distinction only for reader to have a connection to the sensitivity studies shown previously.

\begin{figure}
    \includegraphics[width=\columnwidth]{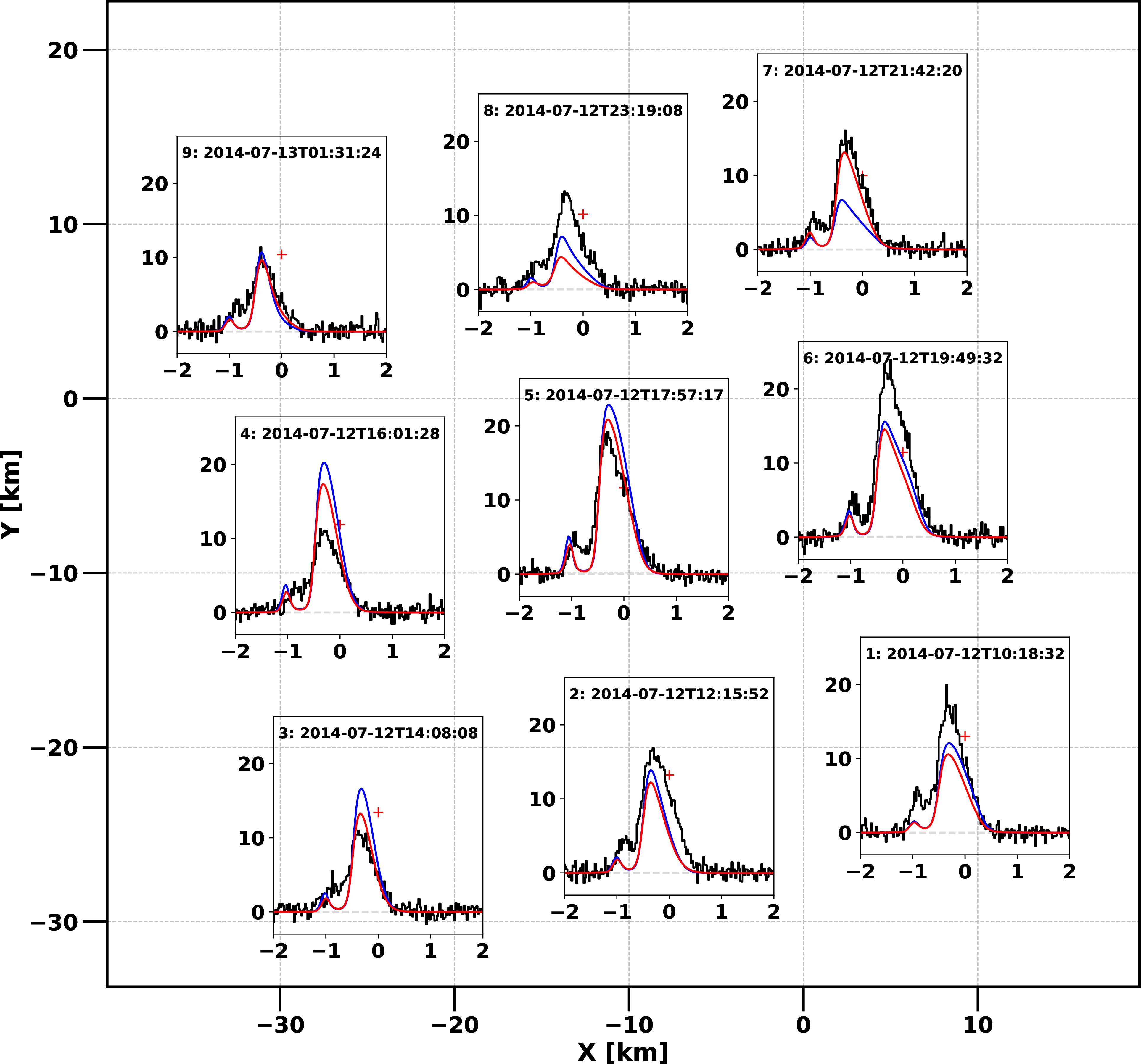}
    \caption{A MIRO 3-by-3 map for July 12, 2014, each inset plot showing the measured spectra (black), Hapi only (Hapi$^{0}$) active region (blue) and (Hapi$^{+}$) (red) [see text for definition]. The geometry was calculated with SPICE kernels and shown here as an offset from the nucleus center projected onto the viewing plane of MIRO (average pointing is displayed as red cross). The inset figures are labeled from 1 through 9 in order in which the pointing was performed. The mid-point time stamp of integration is also shown. The x-axis and y-axis of the inset plots are in units [km/s] and [K] respectively.}
    \label{fig:map1}
\end{figure}

In Fig.~\ref{fig:map1}, we show the full map performed on July 12, 2014, individual map points are numbered (1-9) and respective pointing geometries are shown as an offset from the nucleus center in [km]. The measurements are plotted in black. The largest amplitude is for map point 6 of about 22~K and the smallest amplitude of approximately 9~K for geometry number 3 and 9. 
When we compare the simulated Hapi$^{0}$ (blue) and Hapi$^{+}$ (red) spectra with the observations, we find generally a good agreement in the line shapes, for both the red and the blue wings. In fact the red wings show an excellent fit with observations, exhibiting the correct slopes at the base and core of the lines, which means that the day-night transitions in the coma are captured reasonably well by the model. Nevertheless, the self-absorbed blue wings in the simulated spectra are over-estimated (deeper) than in measurements, more importantly so for the far away pointings (e.g. map points 2, 3, and 9). From the physics of line formation, this would point in the direction of an improper gradient of expansion velocity projected along the MIRO LOS, and perhaps a larger water density as well. In our tests, we were able to modify the coma at individual point to provide a better fit in the blue wing, however, no single correction worked well for all the points in the map. 

Furthermore, the line amplitudes could not be matched precisely for all points in the map with either of the models. The Hapi$^{+}$ is marginally better in agreement with the measured spectra amplitude and also the line width to some degree. However, the fact that calculated line amplitudes at the third pointing is greater than the first is puzzling as it contradicts the measurements. A similar problem appears for points 4 and 6, where calculations do not agree with measurements. From the model point of view, we have checked that the coma generated for time stamps corresponding to the start, middle, or end of the on-source integration do not yield a large difference in line amplitudes. Second, we have rechecked that the pointing stability for the entire integration for a given point (during which 4 instrument calibrations occurred) scatter within about 1-2", which is small compared to 412" or the beam width. Therefore, at this point we cannot adjust the pointing in a way that could improve the fit for this set of observations.

\begin{figure}
    \includegraphics[width=\columnwidth]{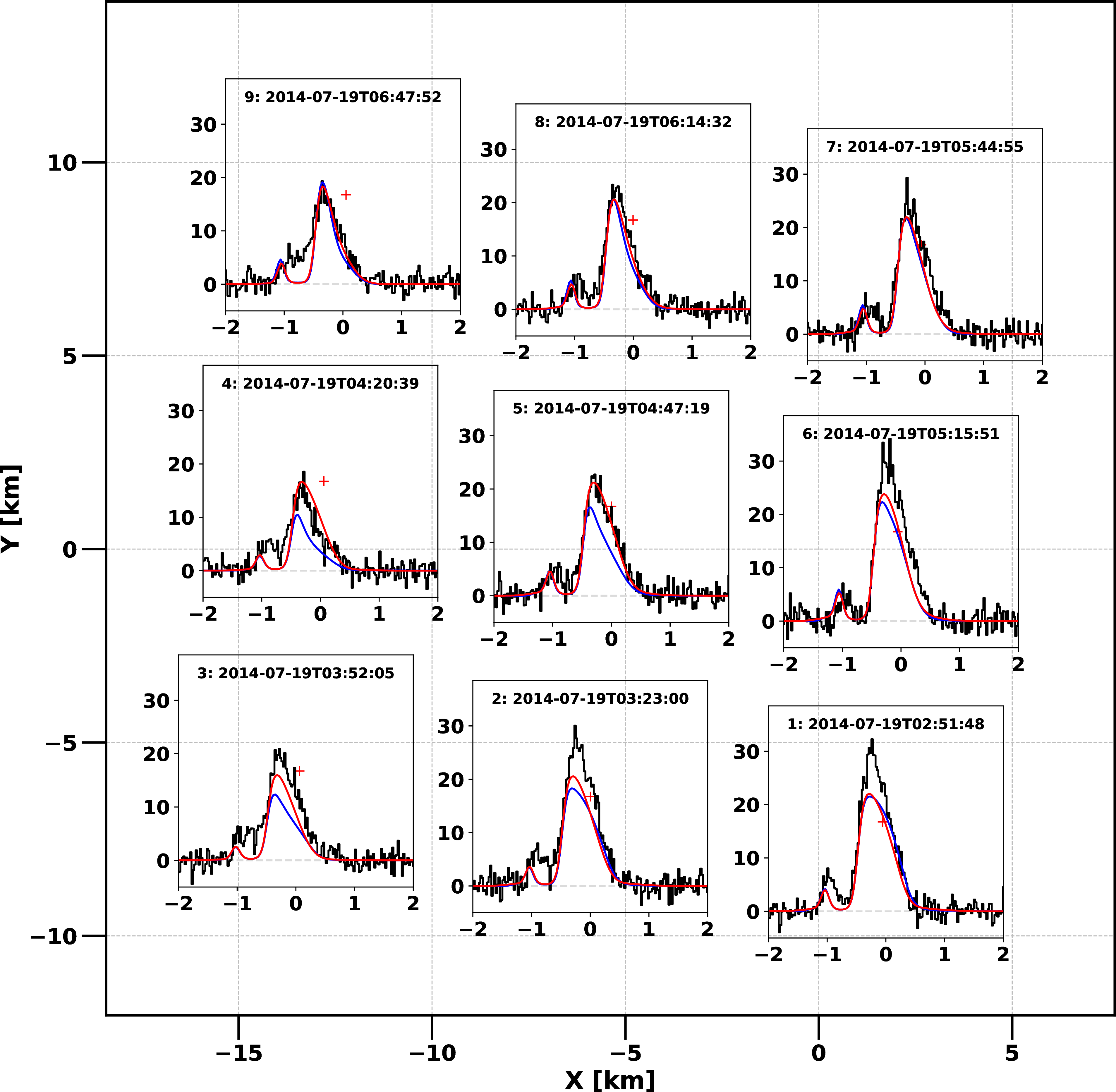}
    \caption{A MIRO 3-by-3 map for July 19, 2014, each inset plot showing the measured spectra (black), Hapi only (Hapi$^{0}$) active region (blue) and (Hapi$^{+}$) (red). The geometry was calculated with SPICE kernels and shown here as an offset from the nucleus center projected onto the viewing plane of MIRO. The inset figures are labeled from 1 through 9 in order in which the pointing was performed. The mid-point time stamp of integration is also shown. The x-axis and y-axis of the inset plots are in units [km/s] and [K] respectively.}
    \label{fig:map2}
\end{figure}

Fig.~\ref{fig:map2} shows another mapping observation we investigated in detail, July 19, 2014. It is also a 3-by-3 mapping campaign but the comet-spacecraft distance was about half compared to that of July 12, 2014.

In this case the entire map can be modeled consistently, without having the unexplained issues observed in July 12, 2014 where the amplitudes did not match observations. For July 19, 2014 the model correctly predicts the line amplitude variations for different geometries, and it is clearly visible that the Hapi$^{+}$ model provides a better overall agreement with the observations. We can also conclude that the line width at the base, the red wing slope, and generally the red wing amplitude is reproduced very well. For pointings 1,2,3 and 6 the peak line amplitudes are still little under-estimated. The model also predicts a deeper self-absorbed core of the line in nearly all cases. As previously noted, there can be different ways this could appear a) projected velocity along the LOS can have different gradients, b) total water density and excitation conditions. Tentatively, a different pointing would also change the behavior in this spectral regions, but it is not clear how to justify this for these observations. 

Similar conclusion follows for the other mapping observations in Table~\ref{tbl:log}. In addition, for later maps we have direct pointing at the nucleus and we begin to observe a portion of the line core in absorption (as the nucleus is warmer than the coma). Although our models do reproduce this effect, we could not find a reasonable fit to the measurements. An extreme case is the map for July 27, 2014, in which we cannot even match the continuum of the lines. It is not clear at this point whether pointing obtained from the SPICE kernels is to blame, or the fact that our nucleus model does not feature even a rudimentary thermal model. It is possible that the issue is a mixture of the two effects, and the advantage of this excellent data MIRO provided during this period can be interpreted with improved physical modeling. This should be a focus of future work.

\section{Summary and perspective}
This work concentrated on the 3D modeling of early MIRO/Rosetta mapping observations, when the nucleus was not yet spatially resolved by MIRO, in attempt to constrain a spatial pattern of H$_{2}$O activity at its onset. We wanted to learn whether 1) the 3D nucleus shape model, proper orientation and illumination driven outgassing can reproduce MIRO line shapes, 2) whether conclusions reached by the 1D modeling applied during the early investigations \citep{Gulkis:2015} will still hold with the 3D model, 3) whether we can identify an extended water source driven by CO$_{2}$ outgassing from the Imhotep region \citep{Hassig:2015}, and if such activity would be consistent with the MIRO observations, and 4) whether we can explicitly model the enhanced Hapi contribution to the early water coma, a consequence of the north-south transport of material \citep{Keller:2015,Keller:2017}. 

We consider this work can still be an exploratory research into 3D coupling of coma structure, non-LTE line transfer and LOS projection effects for MIRO. We developed an empirical, approximate model for coma temperature, expansion velocity and water density in 3D geometry. Each facet's outgassing rate is driven by the illumination, while the temperature and velocity for each facet are determined by the surface equilibrium temperature while the radial profile follows expression similar to those used for 1D model in \citep{Lee:2011} and \citep{Biver:2015}. Comparisons of our semi-analytical model with the nominal DSMC simulations for similar production rates are in a good agreement regarding water density distribution. This gives us confidence that we capture the main general features of the 3D coma driven by the specific nucleus shape.

In this paper we show that the MIRO measured water line amplitude and the line shape itself carry an information that can be used to estimate which nucleus regions contribute to the total number density inside the MIRO beam. We derive the following conclusions from our numerical study of regional sensitivity and fitting the MIRO measurements:
\begin{itemize}
\item We demonstrated that accounting for the 3D shape model, proper orientation, illumination and viewing geometry, the line shapes may deviate strongly from the 1D spherical approximation. Although this fact is not surprising, it is important to demonstrate the limits of 1D modeling already at distances where the beam size is several times of the nucleus diameter. We would expect this deviation to be even stronger for near nucleus observations.
\item The MIRO line shapes show a strong dependence on water activity originating at different nucleus region.  In this connection, we found that the Imhotep outgassing, even if producing enough water molecules, would not be able to match the MIRO spectral line shapes. We conclude that the Imhotep region contributes a small fraction of the total number of water molecules into the MIRO beam due to direct surface outgassing. This is also true for potential extended water source in the coma originating from Imhotep under the assumption that the icy particles follow a radial expansion.
\item We investigated effects of nucleus rotation on the MIRO line shapes considering activity only from the Imhotep, Hapi and also the entire illuminated surface. In these simulations, the variations of the line shape and the line amplitude due to changing insolation as the nucleus rotates are detectable. This is additional evidence that MIRO measurements can constrain regional activity, in qualitative and to some degree qualitative way.
\item Fitting the MIRO measurements with a single 3D coma model (accounting only for changing illumination due to rotation) for all the map points for July 12 and July 19, 2014, we conclude that the Hapi region is indeed the strongest contributor. However, other parts of the nucleus must be active as well in order to provide a good fit to the line widths and amplitudes. We found a reasonable match to observations for scaling the  illumination driven outgassing in other regions by factor 0.1 except the Hapi region, which is scaled up by 1.5 from its nominal illumination value. We also find that there must exist a very strong day/night dichotomy in the water density in the coma during this onset of activity, suggesting only a very little day/night transport of water around the nucleus.
\item Although regional outgassing sensitivity was the main focus of this work, we can also estimate the average total production rate for each map using this model. We evaluate the Q[H$_{2}$O] to be (2.5$\pm$1)$\times 10^{25}$ for July 12, 2014, and (1.8$\pm$0.3)$\times 10^{25}$ for July 19, 2014. The standard deviation represents a scatter of values for different pointing geometries. The uncertainty in total Q[H$_{2}$O] is estimated to be approximately a factor 2 due to uncertainties in coma temperature, and to some degree due to electron density (and temperature) and spectral calibration. The spectroscopic and collisional parameters do not play a strong role by themselves during this period. The derived Q[H$_{2}$O] agree with the early evaluations reported in \citet{Gulkis:2015} using 1D modeling. This agreement is encouraging for 1D spherical calculation being used safely for the total water production rates estimation, since this quantity is related to the line area rather than line shape. However, when fitting the line shape, one has information on the 3D structure of the coma as projected into the LOS of the instrument, and hence can constrain the physics of nucleus activity and/or dynamics in the coma. 
\item The case for constraining the regional nucleus activity is even stronger for later observation as spacecraft near the comet and the projected MIRO beam becomes smaller. In addition, the case of July 27, 2014, provides a more detailed mapping (7x8 and 7x7 points) and the integration times are also smaller. Nevertheless, we find the nominal illumination driven outgassing, simple equilibrium surface temperature, and perhaps even  pointing provided by the SPICE kernels contribute to the fact that we cannot match the line shapes, the line continua,  and ultimately the line amplitudes in any reasonable way. Therefore, these observations should be revisited with more physical models, including proper thermal model of the sub-surface. 
\item Finally, we also note that there is a need for a sharp coma terminator to fit the MIRO line shapes in July 2014. We have achieved this effect by using an opening angle of outgassing for each facet. It is not yet obvious that 3D DSMC codes will be able to reproduce this effect, but it should be investigated in the future.
\end{itemize}

This work constitutes the first attempt to look at the MIRO data using 3D coma, non-LTE radiative transfer model and beam synthesis codes, accounting for the nucleus shape. The Rosetta/MIRO data provide an excellent opportunity to start building and testing physical models of heat and mass transport from subsurface and the relationship to the kinetics  in the coma. We provide first validation that 1D spherical models can be adequately used during this early period for total water production rate estimations (matching line area). However, reproducing the measured line shapes carried additional information which cannot be captured properly with 1D models. The 3D models of the coma can test regional or geomorphological units for activity, and they may include more complex physics of gas flow, or test dust/gas interaction. We do not suggest that MIRO spectra alone cannot uniquely constrain all these parameters, however, we believe that multi-instrument self-consistent modeling should be the way into the future analysis of Rosetta data. Only in this way, we may hope to test validity of the current models of physics of activity and plan for the future observations.

\section*{Acknowledgements}
Rosetta is an European Space Agency (ESA) mission with contributions from its member states and NASA. We acknowledge the entire ESA Rosetta team and thank them for their contribution,  without which this work could not have been done. This extends also to the ESA planetary science archive and the MIRO PI M. Hofstadter (NASA/JPL, Pasadena, USA).  YZ thanks the National Natural Science Foundation of China (Grant No: 11761131008, 11673072, 11633009) and the Foundation of Minor Planets of Purple Mountain Observatory. LR was partially funded from the following grants during this work: Rosetta-MIRO: DLR FKZ 50QP1302, DFG-HA3261-9/1 and the DFG project SPP-1488/2. JJ thanks the National Natural Science Foundation of China (Grant No: 11661161013, 11473073), the Strategic Priority Research Program on Space Science, the Chinese Academy of Sciences, (Grant No. XDA15020302) and the CAS interdisciplinary Innovation Team foundation.







\appendix
\section{The usage of openning angle}
Several tests have been performed varying the opening angle, $\theta$, as defined in section \ref{sec:model}. It was found that using  $\theta$=90$^{\circ}$ (pure hemispheric outgassing from each facet) produces a strongly homogeneous coma, but more importantly the simulated MIRO line shapes implied too many molecules around the terminator. Therefore, we settled to use $\theta$=60$^{\circ}$ in our nominal models, which kept the dayside coma structure, but provided a better defined terminator effect, as contrasted in Fig.~\ref{fig:opa60vs90} below.

It should be understood that $\theta$=60$^{\circ}$ is not a precise required value. There is an entire range of opening angles (55-75$^{\circ}$) which would provide the same conclusions within other uncertainties of the modeling, and the MIRO measurement uncertainties.


 \begin{figure}
	\includegraphics[width=\columnwidth]{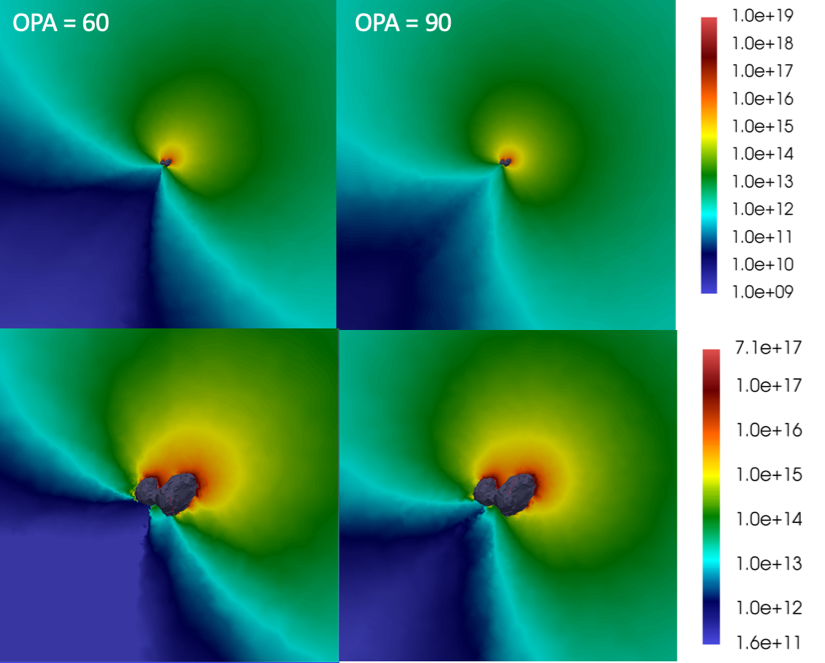}
    \caption{Comparison of the water density $[m^{-3}]$ calculated from our coma model for two opening angle: $\theta = 60^\circ$ for the left two panels and $\theta = 90^\circ$ at the right hand side. The scale of the coma is about 60 km in the upper two panels and about 20 km in the bottom ones. Suitable 3D structure around terminator could not be produced properly when $OPA = 90^\circ$ as shown in the right pannels. The plots show slices when looking into the +Y direction in CG's body frame.}
    \label{fig:opa60vs90}
\end{figure}

\section{MIRO mapping observation July 12, 2014}
\begin{figure}
	\includegraphics[width=\columnwidth]{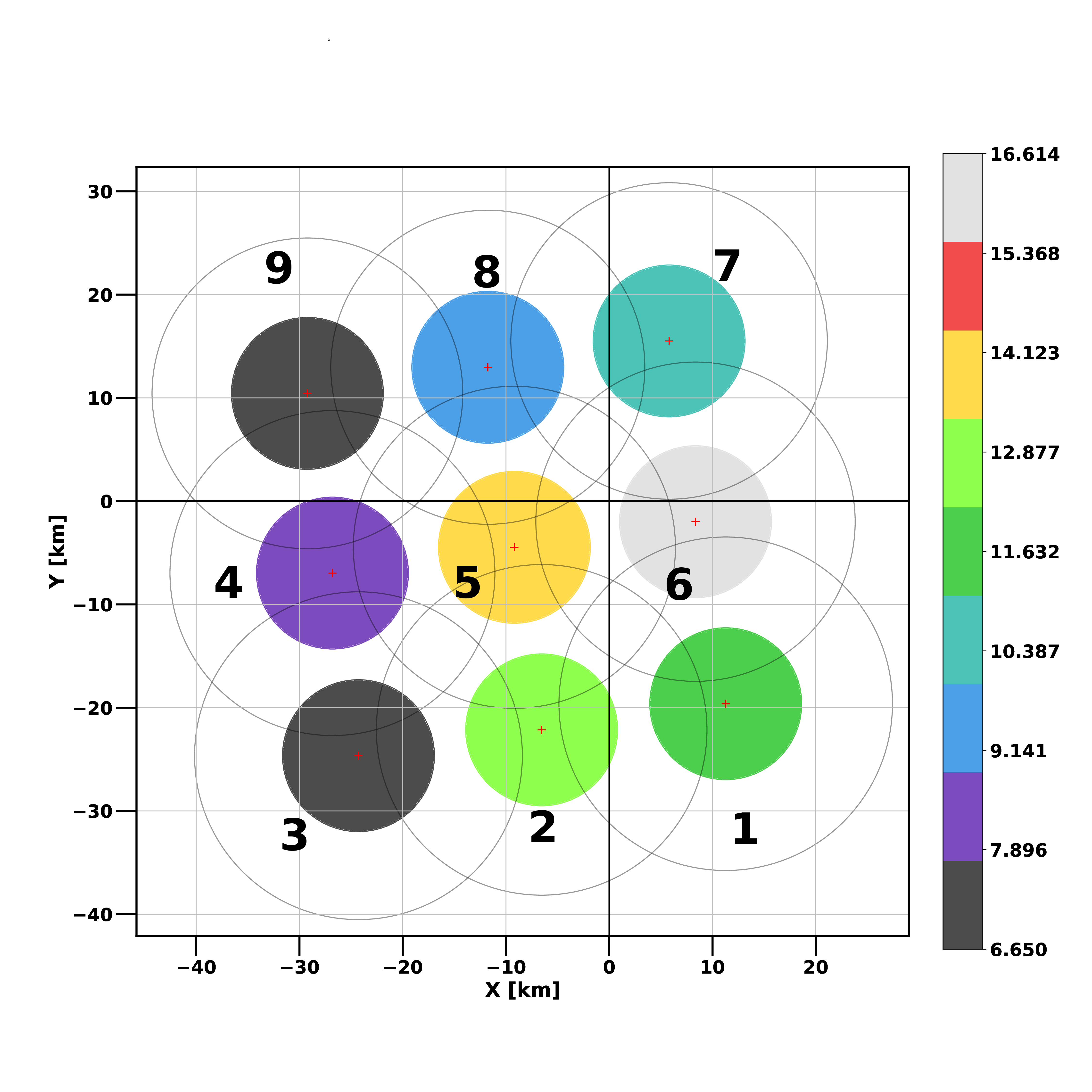}
    \caption{A plot of the MIRO geometry and line areas for the July 12, 2014 map. The x- and y-axis show the pointing distance [km] of MIRO relative to the nucleus center projected onto the viewing plane using the SPICE library. The comet center is at (0,0) on the plot. The average pointing during an integration at a specific point is marked with a red cross, an outline of the beam (FWHM) is plotted as black larger circles, and the color filled circles contains information about the integrated line area [K km/s] (as indicated in the colorbar). The size of the color filled circle is arbitrary, and set only for good visual clarity. The numbers associated with each point in the map give a clue  about the order the observation were made, and we also refer to them in the main text when discussion a particular pointing geometry, e.g. map point number 1, 2, etc.}
    \label{fig:july12map}
\end{figure}

It is very challenging to contain all the necessary information for the interpretation of MIRO observations in a single figure. Especially the mapping observations of MIRO are difficult to visualize with respect to the offset from the comet center, the beam size, the illumination condition of the nucleus, coma temperature, density, expansion velocity and other parameters. In the main text, when discussing an isolated effect of a coma or nucleus on the MIRO line shape, we limit the information only the relevant parameter in question, neglecting others for clarity. Therefore, here we would like to provide, in addition, an overall context of the 3x3 mapping observation in July 12, 2014 (Fig.~\ref{fig:july12map}), which is used in this study.


\end{document}